\documentclass[aps,preprint,amsmath,amssymb,floatfix]{revtex4-1}

\usepackage{graphicx}
\usepackage{dcolumn}
\usepackage{bm}
\usepackage{color}
\usepackage{soul}

\newcommand{\mbf}[1]{\mathbf{ #1 }}

\RequirePackage[normalem]{ulem} 
\RequirePackage{color}\definecolor{RED}{rgb}{1,0,0}\definecolor{BLUE}{rgb}{0,0,1} 
\begin{document}

\title{First-principles simulations of glass-formers}

\author{Walter Kob}
\email[]{walter.kob@umontpellier.fr}
\author{Simona Ispas}

\affiliation{Laboratoire Charles Coulomb, UM-CNRS UMR 5221, 
 Universit\'e Montpellier, Place Eug\`ene Bataillon\\
 F-34095 Montpellier, France }


\begin{abstract}

In this article we review results of computer simulation of glasses
carried out using first principles approaches, notably density functional
theory. We start with a brief introduction to this method and compare
the pros and cons of this approach with the ones of simulations with
classical potentials. This is followed by a discussion of simulation
results of various glass-forming systems that have been obtained via {\it ab
initio} simulations and that demonstrate the usefulness of this approach
to understand the properties of glasses on the microscopic level.

\end{abstract}

\maketitle

\section{Introduction}

In their early days,  i.e. in 1940-1950, computer simulations
were mainly used to address questions concerning the domain of
statistical physics, such as the properties of hard sphere systems or
the dynamics of simple crystals~\cite{frenkel_smit,allen_tildesley}. A few decades
later, i.e. when computers became more powerful and more accessible,
researchers started to use simulations to study the properties of real
materials. For this, people used an approach that today is referred
to as ``classical molecular dynamics'', i.e. the interactions between
the particles that constitute the material are described by an effective
potential with a  form (e.g. Lennard-Jones)  chosen in a rather  {\it
ad hoc} manner. The parameters of the potential (e.g. the depth and the
position of the well in the Lennard-Jones potential) are chosen such
that certain macroscopic properties of the simulated system (such as
its density or the melting temperature of the crystal) match with the
experimental data for the material. Once the interactions are known
one solves numerically Newton's equations of motion and hence
obtains the trajectories of all the particles (for details see Chap.~XX by A.~Takada 
in the present volume~\cite{takada2016encyc}, or Ref.~\cite{frenkel_smit}). From
these trajectories one then can measure the physical properties of
the system, such as the radial distribution function, the diffusion
coefficient, or the mechanical properties such as the elastic constants.

Although simulation studies with effective potentials are very
valuable to gain insight into the qualitative properties of liquids
and solids, they usually do not allow to obtain a good {\it quantitative}
description of a given specific material. The reason for this is that most
material properties depend in a quite sensitive manner on the potential
that describes the interactions between the particles that constitute
the material. Since normally these interactions depend not only on
the type of atoms considered, but also on the microscopic environment
of the particles (e.g. the bond strength between an oxygen atom and a
silicon atom will change if a hydrogen atom is approached to this pair),
it is basically impossible to come up with a simple expression for the
potential energy that is able to describe faithfully all these different
environments. This problem is particularly pronounced for the case of
glasses, since, in contrast to crystals, in these systems one has a
multitude of very different local environments for each atom species.
The only method that at present can address this problem in a systematic
manner is the {\it ab initio} formalism. In this approach one does
not rely on a fixed functional form for the interaction potential, but
instead uses the electronic degrees of freedom of each atom to compute
the force that acts on each atom of the system. Once this force is known,
one can use Newton's equations of motion to determine the trajectory of the
particles and hence subsequently the properties of the system. Thus {\it
a priori} the only input needed for these simulations are the species
of the particles, a feature that has allowed {\it ab initio} simulations to 
become a highly interesting method to gain a detailed understanding
of the properties of glass-forming liquids and glasses. The goal of
this review is to give a brief introduction to the method, discuss its
advantages but also its problems, and then to present some specific examples
that show that this type of simulation is indeed very useful to improve
our understanding of glasses on a microscopic scale. We will conclude
with a brief outlook on what will be possible to do with {\it ab initio}
simulations in the next few years.

\section{Ab Initio Simulations}

In the Introduction we mentioned that within the {\it ab initio} approach
the forces on the atoms are obtained from their electronic degrees of
freedom. The first step to do this is to start with the Schr\"odinger
equation which describes the electronic degrees of freedom of the system
and thus is given by

\begin{equation}
\mathcal{H}_e   \Psi = E  \Psi \quad,
\label{schrodinger}
\end{equation}

\noindent
where the (complex) many-electron wavefunction $\Psi  \left
(\{\mbf{r}_i\}; \{\mathbf R_I\}\right)$ depends on the positions of the
$n$ electrons, $\{\mathbf r_i\}$,  as well as the positions of the $N$
nuclei, $\{\mathbf R_I\}$ \cite{kohanoff2006electronic}. $E$ is the energy of the electronic degrees
of freedom, and the operator $\mathcal H_e$ is given by

\begin{equation}
\mathcal{H}_e =-\sum_{i=1}^{n} \frac{1}{2 }\nabla_i^2
+ \sum_{\stackrel{i,j=1}{j<i}}^{n} \frac{ 1}{|\mbf{r}_{i}-\mbf{r}_{j}|} -
 \sum_{i=1}^{n} \sum_{I=1}^N \frac{Z_I}{|\mbf{r}_{i}-\mbf{R}_{I}|} \quad,
\end{equation}

\noindent
where $Z_I$ is the charge of ion $I$.  Note that this equation is
written in atomic units, i.e.  Planck's constant,  the  electronic
charge  and mass are set to unity, and the Laplacian $\nabla^2$ is
given by $\nabla^2=\partial^2/\partial x^2+ \partial^2/\partial y^2 +
\partial^2/ \partial z^2$. We emphasize that in Eq.~(\ref{schrodinger})
only the electronic degrees of freedom are treated quantum mechanically
whereas the ones of the nuclei are not. This approximation, often
called ``Born-Oppenheimer'' or ``adiabatic'', is quite accurate since
the mass of the electron is almost 2000 times smaller than the one of
the lightest nucleus (i.e. the one of H). Hence the degrees of freedom
of the electrons are basically decoupled from the ones of the nuclei,
i.e. the heavy nuclei move more slowly than the light electrons and
the electrons adapt instantaneously to the changes of the nuclear
positions. As a consequence it can be assumed that for the electronic
structure calculation the nuclei are clamped at fixed positions and
hence the electronic wavefunction $\Psi$ depends on $\{\mathbf R_I\}$
only parametrically~\cite{marx_hutter}.

In order to reduce the computational complexity even further one considers
only the ground state solution of Eq.~(\ref{schrodinger}), $\Psi_0$,
i.e.  the state with the lowest energy. The interaction potential between
the nuclei is then given by  $ \displaystyle \mathbf \Phi (\{\mathbf
R_J\})=  \left\langle \Psi_0 | \mathcal{H}_e |\Psi_0 \right\rangle
+\sum_{I,J=1, J<I}^{N} \frac{Z_I Z_J }{|\mbf{R}_{I}-\mbf{R}_{J}|}$ and
hence the force on particle $I$ is given by $F_I=-\nabla_I \mathbf \Phi
(\{\mathbf R_J\})$~\cite{marx_hutter}.

Finding the wavefunction $\Psi_0$ that is the solution of the
Schr\"odinger equation~(\ref{schrodinger}) is a formidable task
and there are two main approaches to do this: The first one is a
wavefunction-based method, often known as the quantum chemistry
approach, which starts from the Hartree-Fock method and, in a first
order approximation, factorizes the many-body electronic wavefunction
into one-particle wavefunctions. Subsequently one searches numerically
for the ground state of this many-body wavefunction, a task that can be
solved nowadays in a reasonable amount of computer time if the number of atoms
is not too large, say, of order of some tens of atoms.  Nevertheless one
can note some recent works proposing new methods in this field with
very promising results for systems containing some hundreds of atoms
\cite{booth2013towards,eriksen2015linear}.

The second method is the so-called ``density functional
theory'' (DFT), a formalism that exploits certain ground state
properties of a many-electrons system in an external field,
thus in our case the Coulomb field of the nuclei. Following up
ideas proposed by Thomas and Fermi in the 1930's, the current DFT
formalism has been rigorously established by Hohenberg, Kohn, and
Sham~\cite{hohenberg1964inhomogeneous,kohn1965self,kohn1999nobel} in
the early 60's. DFT gets rid of the many-body wavefunction, that depends
on $3 \times n$ electronic spatial coordinates, and replaces it by the
more simple electronic density $\rho(\mbf{r})$ that depends only on
three spatial coordinates:

\begin{equation}
\rho(\mbf{r})= \int \ldots \int d\mbf{r}_2 \ldots d\mbf{r}_n
\Psi_0^*(\mbf{r},\mbf{r}_2,\ldots,\mbf{r}_n)
\Psi_0(\mbf{r},\mbf{r}_2,\ldots,\mbf{r}_n)\label{density_KS}.
\end{equation}

\noindent
Kohn and Sham~\cite{kohn1965self} showed that this density can be written
as a sum of the density of non-interacting particles, $\displaystyle
\rho(\mbf{r}) =  \sum_i^n |\phi_i(\mbf{r})|^2$, where $\phi_i$ are the
fictitious one-particle Kohn-Sham (KS) orbitals, and thus
the ground state density $\rho_0(r)$ is given by the sum of the ground
states of these particles. Hence the total energy of the system can be
expressed as

\begin{equation}
E_{\rm{KS}}[\rho]  =  E_{kin} [\rho]+ \int\mathrm d \mathbf r  \rho ( {\mathbf r})  
\left[ \frac{1}{2}\int d\mathbf r' 
\frac{\rho (\mathbf r')}{|\mathbf r -\mathbf r'|}
 - \sum_I\frac{Z_I}{{|\mbf{r}_{i}-\mbf{R}_{I}|}}\right] +E_{xc}[\rho] .
\label{ks_functional}
\end{equation}

\noindent
Here $E_{kin}[\rho]$ is the kinetic energy of a system of $n$ non-interacting
electrons having the density $\rho$: $E_{kin}=\displaystyle -\frac{1}{2}
\sum_{i=1}^n \int \mathrm d \mathbf r  \phi^*_i(\mbf{r})\nabla^2
\phi_i(\mbf{r})$. The second term represents the Coulomb interactions
between electron-electron and electron-nuclei. The third term is the
so-called exchange-correlation (XC) energy and it accounts for all
quantum many-body effects  due to the Pauli exclusion principle which
introduces correlations between the electrons. Since this term cannot be evaluated exactly,
approximations have to be made. Even if  it is found that $E_{xc}[\rho]$
is usually substantially smaller than those of the two other terms, the
choice of its approximation may become  crucial for chemically complex
systems \cite{jones2015density,burke2012perspective}.  The simplest
one, proposed originally by Kohn and Sham \cite{kohn1965self},  relies
on the assumption that at each point the exchange-correlation energy
density corresponds to the one of a homogeneous gas of electrons. This
approximation is called the ``Local Density Approximation'' (LDA) and
is given by:

\begin{equation}
E_{\rm{xc}}^{\mbox{\scriptsize{LDA}}}[\rho(\mbf{r})]  =  \int d\mbf{r} \rho(\mbf{r})
\varepsilon_{\rm{xc}}^{\mbox{\scriptsize{LDA}}}\left[ \rho(\mbf{r}) \right].
\label{LDA_func}
\end{equation}

We note that even for such a simple model system the expression of the
correlation energy has to be calculated numerically using Monte Carlo
methods. Some more advanced approximations, the so-called ``Generalized
Gradient Approximation'' or GGA, are based on more complex operators
making use of the density gradient of $m$th order

\begin{equation}
E_{\rm{xc}}^{\mbox{\scriptsize{GGA}}}[\rho(\mbf{r})]  =  \int d\mbf{r} \rho(\mbf{r})
\varepsilon_{\rm{xc}}^{\mbox{\scriptsize{GGA}}}\left[ \rho(\mbf{r}); 
\nabla^m \rho(\mbf{r}) \right]
\label{GGA_func}.
\end{equation}

\noindent
However, it is not always the case that these higher approximations give
more reliable results and hence it is {\it a priori} not always clear which
exchange functional should be used~\cite{jones2015density,burke2012perspective}. 
Despite this problem one can say
that the Kohn-Sham approach and reasonable (simple) approximations for
the exchange correlation term have opened the door to the calculations
of the electronic structure for many-atom systems in order to study
real materials.

Although expressing the full quantum mechanical problem in the language
of DFT leads to a significant reduction of the computational effort for
calculating the forces on the nuclei, it is found that in practice this
task is still extremely demanding once one has more than a few tens of
atoms. Therefore one usually makes the further approximation that all
the core electrons of an atom are lumped together and their effect is
replaced by an effective potential, the so-called ``pseudo-potential'',
for the remaining valence electrons which are described by a 
pseudo-wavefunction~\cite{marx_hutter,kohanoff2006electronic}. The
physical motivation for this approximation is that the chemical bonds
between two atoms are usually related to the outer valence electrons
and depend only weakly on the inner core electrons.

So far we have discussed how DFT allows to obtain the forces exerted on the 
nuclei due to their surrounding electrons. These forces can now
be used to solve the equations of motion for the nuclei

\begin{equation}
M_I \ddot{\mbf{R}_I} =  -\nabla_I E_{KS} \left[\{ \phi_i \},\mbf{R} \right] \quad ,
\label{eq_cpmd}
\end{equation}

\noindent 
where the energy $E_{KS}\left[\{ \phi_i \},\mbf{R} \right]$ can
be calculated from the KS orbitals within the Kohn-Sham scheme of
the DFT. In the above equations the nuclei are considered as classical
particles and Eqs.~(\ref{eq_cpmd}) are solved using the same methods
as in classical MD (see Chap. XX  by Takada \cite{takada2016encyc}
or \cite{frenkel_smit,allen_tildesley} for details on how classical
simulations are done). Due to the resulting motion of the ions,
the electronic structure changes and hence one has in principle to
recalculate the total energy of the electronic ground state, a procedure
that remains, despite the DFT approach, computationally very costly. One
possibility to avoid this problem was proposed in a seminal paper by Car
and Parrinello in 1985~\cite{car1985unified}. The idea of that approach,
today called Car-Parrinello Molecular Dynamics (CPMD), is to introduce a
fictive dynamics to the electronic degrees of freedom and thus recast the
quantum mechanical problem into a classical problem with the electronic
wave-functions as new effective degrees of freedom.  Although the
CPMD approach allows to obtain the correct equilibrium properties of
a system, the introduction of the fictive dynamics for the electronic
degrees of freedom makes that the motion of the system in configuration
space is not completely realistic~\cite{marx_hutter}. Despite this
shortcoming, CPMD simulations have been and still are widely used to
study complex systems by means of computer simulations (see for example
Refs.~\cite{berardo2014probing,akola2014structure,bouzid2015origin}).
(On http://www.psi-k.org/codes.shtml one can find the necessary
software to do such simulations.) The mentioned problem can, however,
be avoided within the so-called Born-Oppenheimer molecular dynamics
(BOMD) in which one solves at each time step the electronic problem. This
approach allows thus to give a correct dynamics but this at the cost
of a increased computational load. Only in the last few years the
numerical algorithms have been improved to such an extent that today
it is possible to simulate within BOMD several hundreds particles
~\cite{jakse2014hydrogen,pedesseau2015first1,pedesseau2015first2,plavsienka2015structural}.

The brief description of the {\it ab initio} simulations that we have given so
far should make it clear that in practice the computer code to carry
out such simulations must be extremely optimized in order to keep
the necessary computer time for the simulations within a reasonable
limit. Therefore it is not really advisable that a researcher writes
his/her proper code, in contrast to the situation of simulation with
effective potentials. Instead people use one of the packages that have
been developed over the years and that are very sophisticated and highly
optimized. Various groups use different approaches to maintain and develop
these packages, the most popular ones being Car-Parrinello (CPMD), VASP,
Quantum Espresso, CP2K, Siesta, CASTEP, etc$\ldots$ (see on http://www.psi-k.org/codes.shtml for a more extended list).  Each of these packages has advantages
and disadvantages regarding the scaling of computational effort with
system size, accuracy, ensembles that can be simulated (microcanonical,
canonical, constant pressure, $\ldots$), quantities that can be calculated,
etc. and therefore it is not really possible to say which one is the best.

\begin{table}
\begin{center}
\begin{tabular}{|l|c|c|}
\hline
&{Classical MD} & \ \ \ \ \ \ {\textit{Ab Initio}  MD} \ \ \ \ \ \ \\
\hline
Number of atoms & {1 000 -  500 000 } &  {100 - 500 } \\
Box size &{$\sim$ 100~\AA} & {$\sim$ 15-20\AA} \\
Trajectory length& {$\sim$ 1~ns $-$ 10~$\mu$s} &  {$\sim$ 20 - 100 ps} \\
Transferability & sometimes & yes\\
\hline
\end{tabular}
\caption{
Comparison of various features of large scale computer simulations carried out with classical and 
{\it ab initio} methods.}
\label{table_1}
\end{center}
\end{table}

In order to give an idea on what at present can be done with such {\it ab
initio} simulations we present in Table~\ref{table_1} a brief comparison
between {\it ab initio} simulations and classical simulations. Since the
computational load for the {\it ab initio} simulations does depend in
a non-negligible manner on the system considered (due to the different
electronic structure for the atoms), we consider the example of a glass
containing oxide.  We note that the above numbers are valid for a certain
amount of computer time and of course one has the choice to trade the
number of atoms versus the time span covered in the simulation. For
a classical system the relevant number is basically the product
of the two quantities, i.e. doubling the system size will increase
the necessary computer time by a factor close to two. For {\it ab initio}
simulation the situation is not that favorable, since doubling of the
system size usually leads to an increase of the computational load by a
factor of $2^\alpha$, with $\alpha$ on the order of 3. As a consequence
{\it ab initio} simulations do not only have smaller system sizes but
also the accessible time scale is smaller than the ones in classical
simulations. Therefore the quench rates that are used to cool a system
from its liquid state to its glass state are usually on the order of
$10^{13}-10^{15}$K/s, rates that are thus significantly larger than the
ones used in classical simulations ($10^{10}-10^{14}$K/s) and of course
much larger than the experimental rates ($10^{-2}-10^6$K/s).  However,
despite these huge differences in cooling rates, the resulting glasses
are surprisingly similar, since many of their properties depend only in
a logarithmic way on these rates. Therefore it does make sense to use
{\it ab initio} simulations to investigate the properties of glasses on
the microscopic scale.

\section{Structural properties}
  
The goal of this section is to present some examples that demonstrate how
{\it ab initio} simulations can help to gain a better understanding of the
local structure of complex glasses. Since the literature on this topic is
already quite substantial, it is not possible to discuss all the results
that have been obtained so far, but we will mention at least some of the 
relevant ones.

The very first {\it ab initio} MD simulations for
a glass-former were carried out by Sarnthein {\it et
al.} in 1995 who considered the archetypical glass-former
silica~\cite{sarnthein1995model,sarnthein1995structural}. Using the
Car-Parrinello approach, they generated a glass model by equilibrating
for 10~ps a liquid sample with 72 atoms (!) at 3500~K and subsequently
quenching it to $T=0$~K. Despite the smallness of the system and the
high quench rate ($10^{15}$K/s!), the resulting glass structure was
surprisingly similar to the one of the real material in that, e.g.,
the neutron structure factor was compatible with the one obtained from
scattering experiments. As expected, the network in the glass was built
from SiO$_4$ corner-sharing tetrahedra. These authors calculated also the
electronic density of states and found it to match well with the one obtained
from x-ray photoelectron spectroscopy experiments, although the predicted
band gap of 5.6~eV underestimated the experimental value of about 9~eV,
a flaw that often occurs in DFT calculations \cite{RMartin_book}. In these
papers also the vibrational properties of the glass were determined and
below we will come back to these results.

This pioneering paper was followed up by further studies in which more
complex glass-formers were studied, such as alkali silicate glasses, 
calcium-alumino-silicates~\cite{benoit2000model,benoit2001structural,ispas2001structural,benoit2001structural,
donadio2004photoelasticity,pohlmann2004first,du2006structure,vuilleumier2009computer,jakse2012interplay,spiekermann2013vibrational,vuilleumier2015carbon}, and other glass formers 
\cite{giacomazzi2005medium,ohmura2008mechanism,ferlat2008boroxol,ohmura2009anomalous}.
These investigations
allowed to obtain detailed insight into the local arrangement of the
atoms, how network modifiers like Na or Ca modify these arrangements,
to connect the local structure in real space with the structural
features as determined in neutron or x-ray scattering experiments,
as well the vibrational features discussed in the next section.

As an example on what type of structural information present day
{\it ab initio} simulations can provide we will now briefly discuss
some results obtained for a sodium borosilicate of composition
30\%~Na$_2$O-10\%~B$_2$O$_3$-60\%~SiO$_2$(NBS), a glass-former that
is the basis of many glasses found in our daily life. In addition this
system is interesting since the boron atoms can form bonds with three or
four oxygen neighbors, which makes that the structure of this system
is rather complex~\cite{pedesseau2015first1,pedesseau2015first2}.
One of the most important quantities to characterize the structure are
the partial radial distribution functions $g_{\alpha\beta}(r)$ which
are directly proportional to the probability that two atoms of type
$\alpha$ and $\beta$ are found at a distance $r$ from each other. Thus
this function is defined as~\cite{hansen_mcdonald,binder_kob}

\begin{equation}
g_{\alpha \beta}(r ) = \frac{V}{4\pi r^2 N_\alpha (N_\beta - \delta_{\alpha \beta})}
\sum_{i=1}^{N_\alpha} \sum_{j=1}^{N_\beta} 
\left \langle \delta(r- |\vec r_i -\vec r_j|) \right \rangle
\quad,
\label{eq_gr}
\end{equation}

\noindent
where $\langle .\rangle$ represents the thermal average, $V$ is the
volume of the simulation box, $N_\alpha$ is the number of particles of
species $\alpha$, and $\delta_{\alpha\beta}$ is the Kronecker delta. 

\begin{figure}
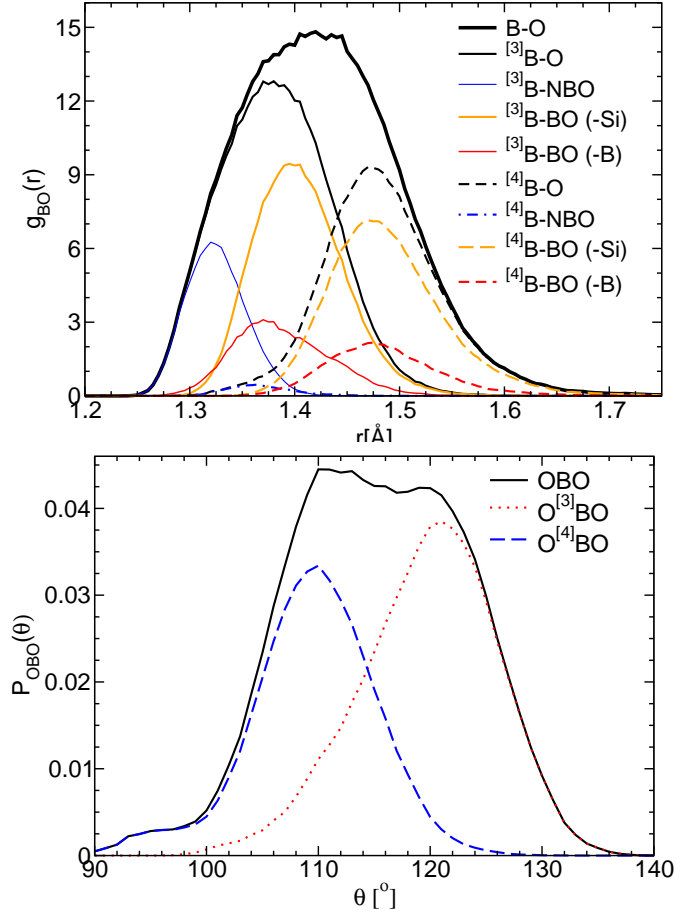

\includegraphics[scale=0.3]{fig-nbs-300K-decomposition-1stpeak-bo.eps}
\includegraphics[scale=0.3]{fig-nbs-300K-decomposition-bad-obo.eps}
\caption{
Top: First peak of the partial radial distribution function $g_{\rm BO}(r)$
of the NBS glass. The total peak has been decomposed
into the various contributions from the different environments around a boron atom.
Bottom: Distribution of the angle formed by two oxygen atoms that are connected
to the same boron atom. This distribution is decomposed into contributions
in which the central boron atoms is of type $^{[3]}$B or type $^{[4]}$B
(dashed and dotted line, respectively). From Ref.~\cite{pedesseau2015first2}.}
\label{fig_nbs_str}
\end{figure}

In the upper panel of Fig.~\ref{fig_nbs_str} we show the radial
distribution function for the boron-oxygen pair, focusing on the
first nearest neighbor peak.  We see that this peak (bold solid black
curve) is relatively large and slightly asymmetric. By considering
the two principal environments of a boron atom (three and four fold
coordinated boron atoms, $^{[3]}$B and $^{[4]}$B, respectively), we can
decompose this peak into two contributions (thin solid and dashed lines,
respectively). This allows us to understand that the broadening of the
total peak is a consequence of the presence of these two populations,
with the bond distance $^{[3]}$B$-$O giving rise to the smaller distances
and the $^{[4]}$B$-$O bonds to the large distances. The distributions for
these two distances can be further decomposed by considering the nature
of the second nearest neighbor of the central boron atoms, i.e. whether
the nearest oxygen is connected to a Si or B atom, or whether it is a
non-bridging oxygen. We see that for the case of the $^{[3]}$B atoms,
the nature of this second nearest neighbor species influences the
B$-$O distance since the position of the corresponding peak depends
on the species, whereas this is not the case for the length of the
$^{[4]}$B$-$O bonds.

In the lower panel of Fig.~\ref{fig_nbs_str} we show the distribution
function for the central angle formed by an O$-$B$-$O triplet (bold black
line). One sees that this distribution has two peaks and again they can
be traced back to the presence of $^{[3]}$B and $^{[4]}$B structural units
(dashed and dotted lines, respectively).

At this point we emphasize that all these structural details can only
be obtained because the structure obtained from the {\it ab initio}
simulations is, at least locally, very accurate. Although one can of
course make the same type of analysis for structures that have been
obtained from a simulation with an effective force field, it is quite
unlikely that an effective potential is able to reproduce the correct
physics (charge transfer between the atoms, $\ldots$) involved in
the formation of these local structures corresponding to different
coordination states.  In contrast to this, one expects quantities
like the positions of the peak and the sub-peaks, their heights or
broadening etc.~will be correctly reproduced by an {\it ab initio}
simulations. Thus for this type of details and especially for glasses
with complex compositions the use of {\it ab initio} simulations
is mandatory. We also mention that for chalcogenide glasses, there are at
present no force-fields available that are able to describe correctly
the homopolar bonds.

\begin{figure}
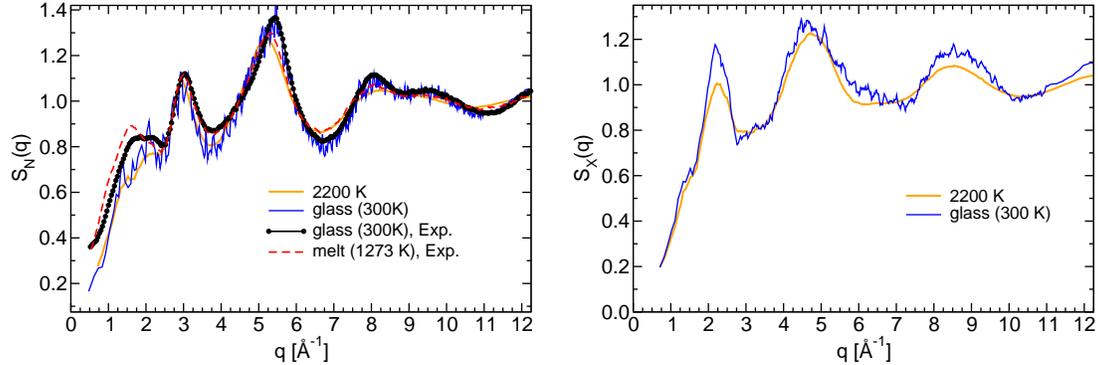

\includegraphics[scale=0.27]{fig1_7-struct-fact-neutron_LIQUID+GLASS-pedesseau2015-2liq-temp-adapted.eps}
\quad
\includegraphics[scale=0.27]{fig1_8-struct-fact-Xray_LIQUID+GLASS-pedesseau2015-1-3liq-temp-adapted.eps}
\caption{Neutron and x-ray structural factors (left and right panel,
respectively) for a borosilicate system in the liquid and glass
state. Adapted from Ref.~\protect\cite{pedesseau2015first1}.}
\label{fig_nbs_str_q}
\end{figure}

The detailed structural information shown in Fig.~\ref{fig_nbs_str} is
very valuable to obtain a better understanding of the local structure of
the glass. It is, however, also important to connect the results from
the simulation with experimental data. Although experiments on atomic
systems do not allow to access directly the radial distribution functions,
it is possible to measure the static structure factor which is directly
related to the weighted sum of $S_{\alpha\beta}(q)$, the space Fourier
transform of $g_{\alpha\beta}(r)$~\cite{binder_kob}. For the case of
neutron scattering the connection is given by

\begin{equation}
S_{\rm N}(q)= \frac{N}{\sum_{\alpha} N_\alpha b_\alpha^2}
\sum_{\alpha, \beta}  b_\alpha b_\beta S_{\alpha \beta} (q) \quad .
\end{equation}

\noindent
Here $b_\alpha$ are the neutron scattering length~\cite{lovesey1984theory} and the
partial static structure factors are given by

\begin{equation}
S_{\alpha\beta}(q) =  \frac{f_{\alpha\beta}}{N}
\sum_{j=1}^{N_\alpha} \sum_{k=1}^{N_\beta}
\left \langle \exp(i{\bf q}.({\bf r}_j - {\bf r}_k)) \right \rangle 
\quad ,
\end{equation}

\noindent
where $f_{\alpha\beta}=1$ for ${\rm \alpha=\beta}$ and
$f_{\alpha\beta}=1/2$ otherwise and $N$ is the  total number of atoms.
For the case of x-ray scattering the relation is

\begin{equation}
S_{\rm X}(q)= \frac{N}{\sum_\alpha N_\alpha f^2_\alpha (q/4\pi)}
\sum_{\alpha, \beta}  f_\alpha (q/4\pi) f_\beta (q/4\pi)S_{\alpha\beta} (q)\, ,
\end{equation}

\noindent
where $f_\alpha (s)$ is the scattering-factor function (also called form
factor) and its value can be found in Ref.~\cite{waasmaier1995new}. For the case of NBS the
$q-$dependence of $S_{\rm N}(q)$ and $S_{\rm X}(q)$ is shown in
Fig.~\ref{fig_nbs_str_q} and we recognize that the two functions are
rather different despite the fact that they have been obtained from
exactly the same glass sample. The reason for this is that these structure
factors are a {\it weighted} sum of the partial structure factors,
each of which has a multitude of positive and negative peaks. Since the
weight depends on whether one considers neutron or x-ray scattering, the
resulting sum might show a peak for, say, $S_{\rm N}(q)$ whereas this feature is
basically absent in $S_{\rm X}(q)$. Since in systems with $k$ species
one has $k(k+1)/2$ partial structure factors, the interpretation of the
various peaks in experimental data can become rather difficult. It is in
such cases that the structure factor as obtained from computer simulations
can help to get the correct interpretation of the various peaks, but this
only under the condition that the positions and height of the different
peaks in the partial structure factors are reproduced reliably, and this
is usually only the case with {\it ab initio} simulations.

\begin{figure}
\includegraphics[scale=0.17]{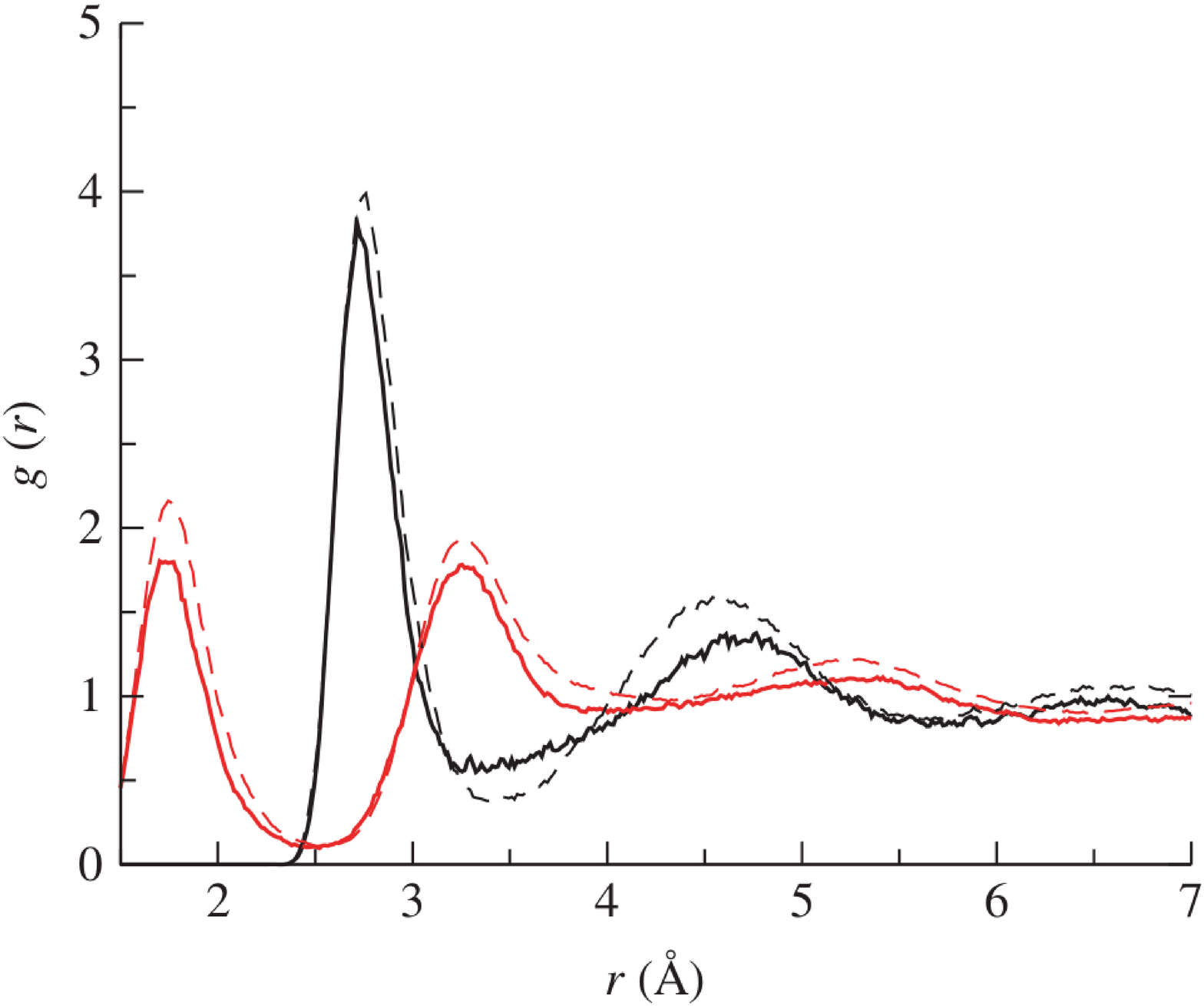}\quad
\includegraphics[scale=0.17]{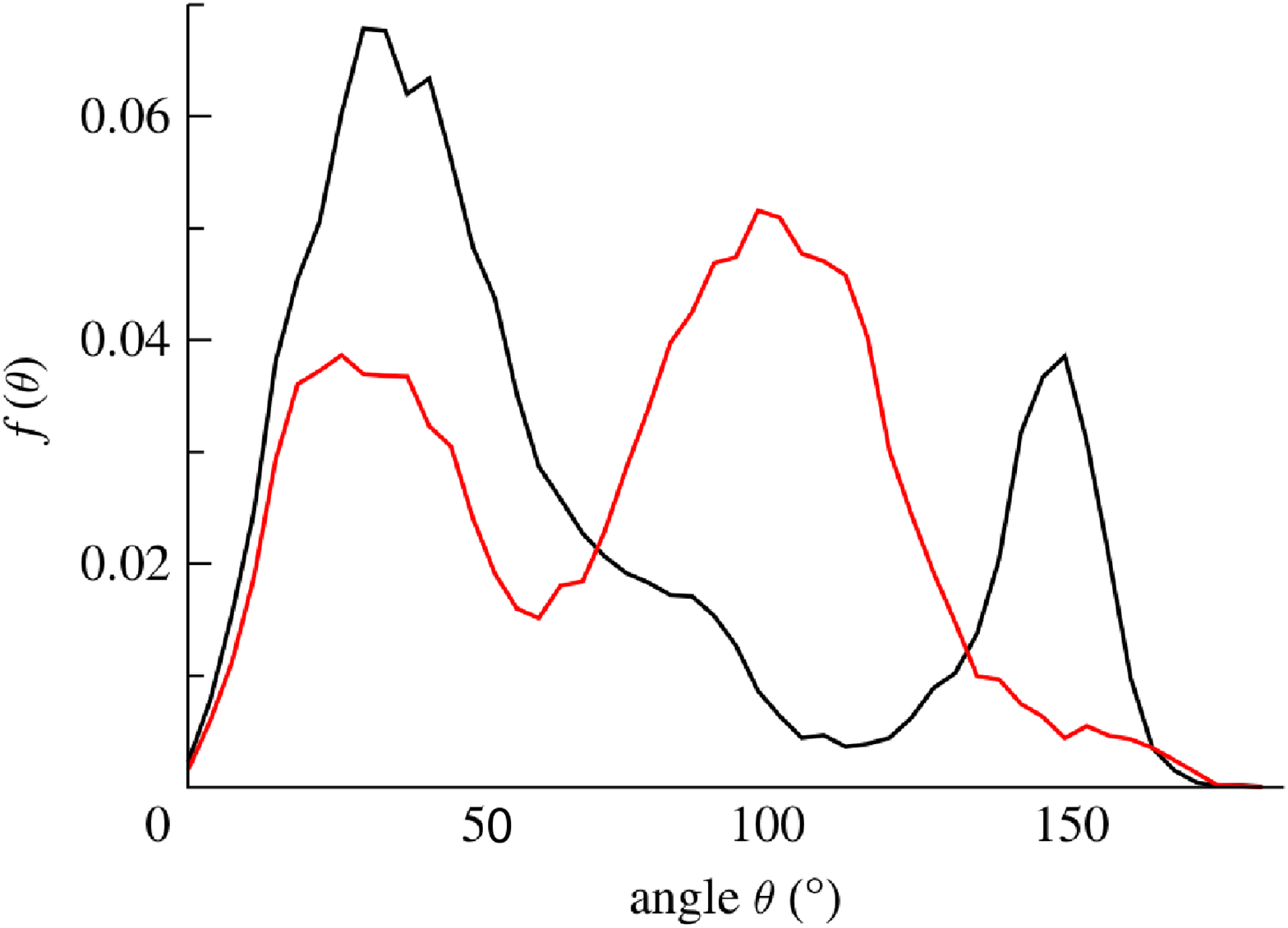}
\caption{ Left panel: Water-water radial distribution functions for
surface water (SW) and bulk water (BW): black continuous line, O-O (SW);
red continuous line, O-H (SW); black dashed line, O-O (BW); red dashed
line, O-H (BW).  Right panel: Distribution of the angle $\theta$  between
the water dipole (bisecting the HOH angle) and the surface normal, for
SW and BW molecules: black solid line, SW; red solid line, BW. Figures
extracted from Ref.~\protect\cite{tilocca2011initial}.}
\label{fig_sio2_h2o_surf} 
\end{figure}

So far we have considered simulations of bulk systems. Another important
situation in which simulations can help to gain a better understanding of
the structure are surfaces, since in this case experimental scattering
techniques are much less powerful to obtain microscopic information
since one has to use grazing ray geometries. Although in principle it is
no problem to use simulations to investigate surfaces, in such studies
one is faced with the issue that basically all effective potentials have
been developed for {\it bulk} systems and there is no guarantee that they
will be reliable for surfaces in which the local structures can be very
different from the ones in the bulk (e.g.~for the case of pure silica one
finds at the surface an appreciable concentration of two membered closed
rings, a structure that is completely absent in the bulk). Hence for such
simulations it is strongly preferable to use {\it ab initio} simulations
since they allow without adjustments of any parameter to deal with such
heterogeneous geometries. As an example for such a simulation we show
in the left panel of Fig.~\ref{fig_sio2_h2o_surf} the partial radial
distribution functions for H$-$O and O$-$O of water that is close to a
surface of amorphous silica (full lines)~\cite{tilocca2011initial}. For
the sake of comparison the graph also includes the corresponding
distributions for bulk water (dashed lines). The curves clearly show that
there is no significant difference between bulk and surface water with
respect to the position of the various peaks. However, for the water close
to the surface, the height of the peaks is lower than the one for the bulk
system thus showing that close to the surface the fluid is less structured
than in the bulk.  Since these modifications are not very pronounced,
it is important to study such effects using interaction potentials that
are very reliable, and thus one has to use {\it ab initio} simulations.
In the right panel of Fig.~\ref{fig_sio2_h2o_surf} we show the angle
between water dipole (i.e. the bissector of the HOH angle) and the surface
normal for the water in the bulk state  as well for water close to the
surface and we recognize that these distributions are very different, thus
showing that the presence of the surface leads to a strong change of the
local geometry. To conclude this section on the structure of glasses we
mention that {\it ab initio} simulations are not only very useful, read
mandatory, if the system has many different components, but also if the
glass contains water~\cite{pohlmann2004first}, an molecule that due to its
reactivity is very difficult to simulate by means of effective potentials.

\section{Vibrational properties}

The structural features of glasses are important for their mechanical
properties as well as for some transport properties (e.g.~in ion conducting
glasses). On the other hand there are many quantities that depend directly
or indirectly on the vibrational properties of the system, such as the
specific heat, the conduction of heat, the transparency of the material,
etc. Therefore it is of great interest to study also these vibrational
properties in detail.

We remind the reader that the vibrational density of states (vDOS)$g(\omega)$,
 is the normalized distribution function of the eigenfrequencies
$\omega_p$ of the dynamical matrix of the system (which in turn is
directly related to the second derivative of the potential with respect
to the coordinates $\vec{r}_i$ and $\vec{r}_j$, i.e.~$\partial^2 
U_{pot}/\partial \vec{r}_i \partial \vec{r}_j$), thus

\begin{equation}
g(\omega) =\frac{1}{3N-3} \sum_{p=4}^{3N} \delta(\omega- \omega_p) \quad .
\label{eq_vdos}
\end{equation}

\noindent
Here one divides by $3N-3$, the number of eigenmodes with non-zero
frequency.  Associated with each eigenfrequency $\omega_p$ of the
dynamical matrix is an eigenvector ${\bf e}_p$ that gives detailed
information regarding the particles that oscillate with the frequency
$\omega_p$. Such studies have allowed to gain insight into the nature
of the vibrational modes of various materials such as silica and germania
glasses~\cite{sarnthein1997origin,pasquarello1998dynamic,benoit2002vibrational,giacomazzi2005medium,giacomazzi2009medium,haworth2010probing}
as well as more complex
systems~\cite{ispas2005vibrational,du2006structure,ganster2007structural,corno2008b3lyp,spiekermann2013vibrational,pedesseau2015first2}.

Note that the true vDOS given by Eq.~(\ref{eq_vdos}) is not directly
accessible in experiments, in which only an  effective vDOS:
$G(\omega) = C(\omega)g(\omega)$ can be measured.   The correction
function $C(\omega)$ describes how the modes with frequency
$\omega$ are coupled to the probing beam of the experiment (photons,
neutrons,...)~\cite{dove1993introduction}. For the case of  inelastic
neutron scattering experiments, this function can be calculated,
within approximations, from the eigenvectors ${\bf e}_p$ of the dynamical
matrix~\cite{taraskin1997connection}.

In practice there are two possibilities to calculate the vibrational
properties of a system. The first one is to determine the local potential
minimum of the glass structure (i.e. to quench the sample to $T=0$~K)
and then to calculate directly the dynamical matrix, e.g.~by using a
finite difference scheme for the forces. By diagonalizing this matrix
one can then obtain its eigenfrequencies and eigenvectors and hence
$g(\omega)$.  The second method consists in running a simulation in
which one solves Newton's equations of motion, and one measures the time
auto-correlation function of the  velocity of a particle.  It can be
shown that the time Fourier transform of this velocity auto-correlation
is directly proportional to the vibrational density of states at low
temperatures, i.e. in a regime in which the harmonic approximation
is valid~\cite{dove1993introduction}.  For higher temperature, a
density of state computed using this approach can present modifications
with respect to the one at low temperatures, which can originate from
anharmonic effects.  We note that the approach with the auto-correlation
function is from the computational point of view inexpensive, but it has
the drawback that it does not give any information on the eigenvectors,
i.e. the nature of the motion of the atoms.

In the previous section we have argued that often {\it ab initio}
simulations are mandatory to get a good qualitative description of the
local atomistic structure. Experience shows that this is even more the
case if one wants to study the vibrational properties of a glass. Roughly
speaking the reason for this is that the structure is related to a balance
of the forces between the particles, i.e. the {\it first} derivative
of the potential, since these forces have basically to compensate in
order to get a mechanically stable structure.  In contrast to this the
vibrational properties are related to the curvature of the potential,
i.e. its {\it second} derivatives. As a consequence it is in practice
quite hard to find effective potentials that are able to reproduce the
experimental vibrational density of states and in fact this is the case
even for systems as simple as pure silica~\cite{huang2015challenges}. 

\begin{figure}
\includegraphics[scale=0.4]{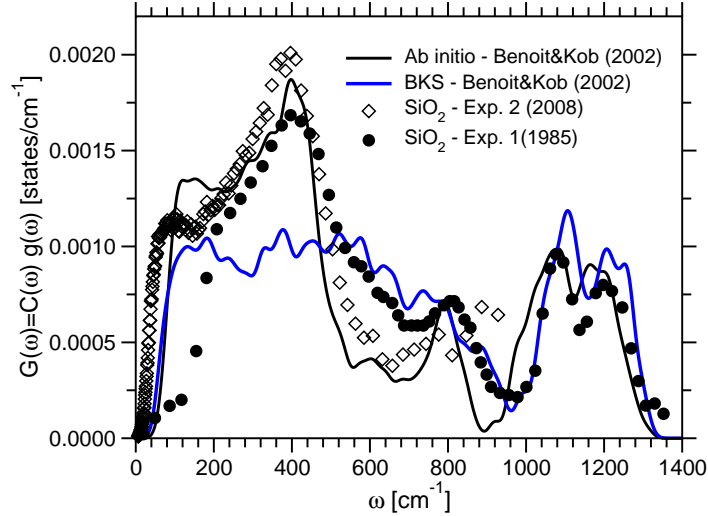}
\caption{Effective neutron densities of states (vDOS) for
a-SiO2 as obtained from an {\it ab initio} simulation (solid
black line)  and a simulation with an effective force field
(BKS, solid blue line)~\cite{benoit2002vibrational}. Also
included is data from two type of neutron scattering experiments
(symbols)~\cite{carpenter1985correlated,fabiani2008neutron}.}
\label{figure-vdos-sio2-benoit-kob02+exp}
\end{figure}

As an example for this we show in 
Fig.~\ref{figure-vdos-sio2-benoit-kob02+exp} the neutron effective vDOS
for silica as it has been obtained by means of a simulation
with the effective potential proposed by van Beest, Kramers,
and van Santen~\cite{vanbeest1990force} which is able to give
a surprisingly good description of the structural properties of
real silica~\cite{vollmayr1996cooling}. Also included in the figure is the {\it
ab initio} data from a small sample of SiO$_2$ (78 atoms)~\cite{benoit2002vibrational}. 
We see that the two predictions are quite similar regarding the vDOS
at high frequencies, i.e. the modes concerning intra-tetrahedral
excitations~\cite{sarnthein1997origin}. However, for frequencies between 200
and 600~cm$^{-1}$ the two calculated spectra differ significantly in that the {\it
ab initio} data shows a marked peak at around 400~cm$^{-1}$ whereas
the BKS data is basically flat. A comparison with the experimental
data (symbols) shows that there is indeed a peak in that frequency range,
i.e.~that the prediction of the {\it ab initio} simulation is correct,
and this despite the smallness of the sample.  Thus we can conclude
that although from the point of view of structure the BKS potential is
quite reliable, it fails to reproduce some of the vibrational features
found in real silica. More details on this point can be found in
Ref.~\cite{benoit2002vibrational}.

A further technique that is used in experiments to get insight
into the vibrational properties of glasses are dielectric
measurements (see Chap.~XX in the present volume by G.~Henderson~\cite{henderson2016encyc}). 
Since this quantity is directly related to the
local polarizability of the material, it is mandatory that the simulation
does give a good description of this polarizability and for the {\it
ab initio} approach this is indeed the case~\cite{gonze1997dynamical}
thus allowing to access directly the high-frequency and static
dielectric constant. The previous, $\epsilon_\infty$, can be estimated
as one third of the trace of the purely electronic dielectric tensor
$(\epsilon_\infty)_{ij}=\delta_{ij}+\displaystyle\frac{4\pi}{V}\frac{\partial^2
E_{tot}}{\partial \mathcal E_i \partial \mathcal E_j}$ (also called
relative permittivity), a tensor that describes the reaction of the
electrons to the presence of an external electric field under the
condition that the ions are kept at fixed positions. In practice it is
found that for glasses this tensor is basically isotropic and diagonal.
The static dielectric constant $\epsilon_0$ reflects the ionic
displacement contributions to the dielectric constant and it can be expressed
as \cite{pasquarello1997dynamical}:

\begin{equation}
  \label{epsilon0}
\epsilon_0 =\epsilon_\infty +
            \frac{4\pi}{3V} \sum_p \sum_j \frac{|\mathcal F_j^p|^2}{\omega_p} \quad ,
\end{equation}

\noindent
where $p$ runs over all the eigenmodes, $j \in\{x, y, z\}$, and $\mathcal
F^{p}_j$ is the so-called oscillator strength which is defined as

\begin{equation}
\mathcal F_{j}^{p} =
\sum_{I,k}
\frac{{\mathbf e_{I,k} (\omega_p)}}{\sqrt{m_{I}}} Z_{I,jk}  \, .
\end{equation}

\noindent
Here $\mathbf e_{I,k}(\omega_p)$ is that part of the eigenvector $\mathbf
e(\omega_p)$ that contains the 3 components related to the displacement
of particle $I$, and the quantity $Z_{I,jk}$ is the Born effective charge
tensor $Z_{I,jk}$ which is given by

\begin{equation}
Z_{I,ij}=\frac{\partial F^{I}_i}{e\partial \mathcal E_j}
  \quad  I=1,2, \ldots , N,   \quad  i, j \in\{x, y, z\} ,
\label{Z_ij}
\end{equation}

\noindent
where $e$ is the elementary charge, i.e. $Z_{I,ij}$ is an effective charge
that connects the strength of an external electric field $\mathcal E$
to the force $\mathbf{F}^{I}$ acting on particle $I$.

From these quantities one can now obtain immediately
the real and imaginary parts of the dielectric function
$\epsilon(\omega)=\epsilon_1(\omega)+i\epsilon_2(\omega)$
\cite{thorpe1986coulomb,pasquarello1997dynamical}:

\begin{eqnarray}
\epsilon_{1}(\omega) &= &
\epsilon_{\infty}-\frac{4\pi}{3V}\sum_{p} \sum_j
\frac{\mid \mathcal F_j^{p}\mid^2}{\omega^2-\omega_{p}^2}  \\
\epsilon_{2}(\omega) &= &
\frac{4\pi^2}{3V}\sum_{p} \sum_j
\frac{\mid \mathcal F_j^{p}\mid^2}{2\omega_{p}^2} \delta(\omega-\omega_{n}).
\end{eqnarray}

Closely related to $\epsilon(\omega)$ is the absorption spectra
$\alpha(\omega)$ which is given by~\cite{kamitsos1994vibrational}

\begin{equation}
\alpha(\omega) = 4\pi\omega n''(\omega)  \quad , \mbox{ with } 
n''(\omega) = \sqrt{\frac{\sqrt{\epsilon_1^2+ \epsilon_2^2}- \epsilon_1}{2}}\quad .
\label{eq_absorption_1}
\end{equation}

\begin{figure}
\includegraphics[scale=0.25]{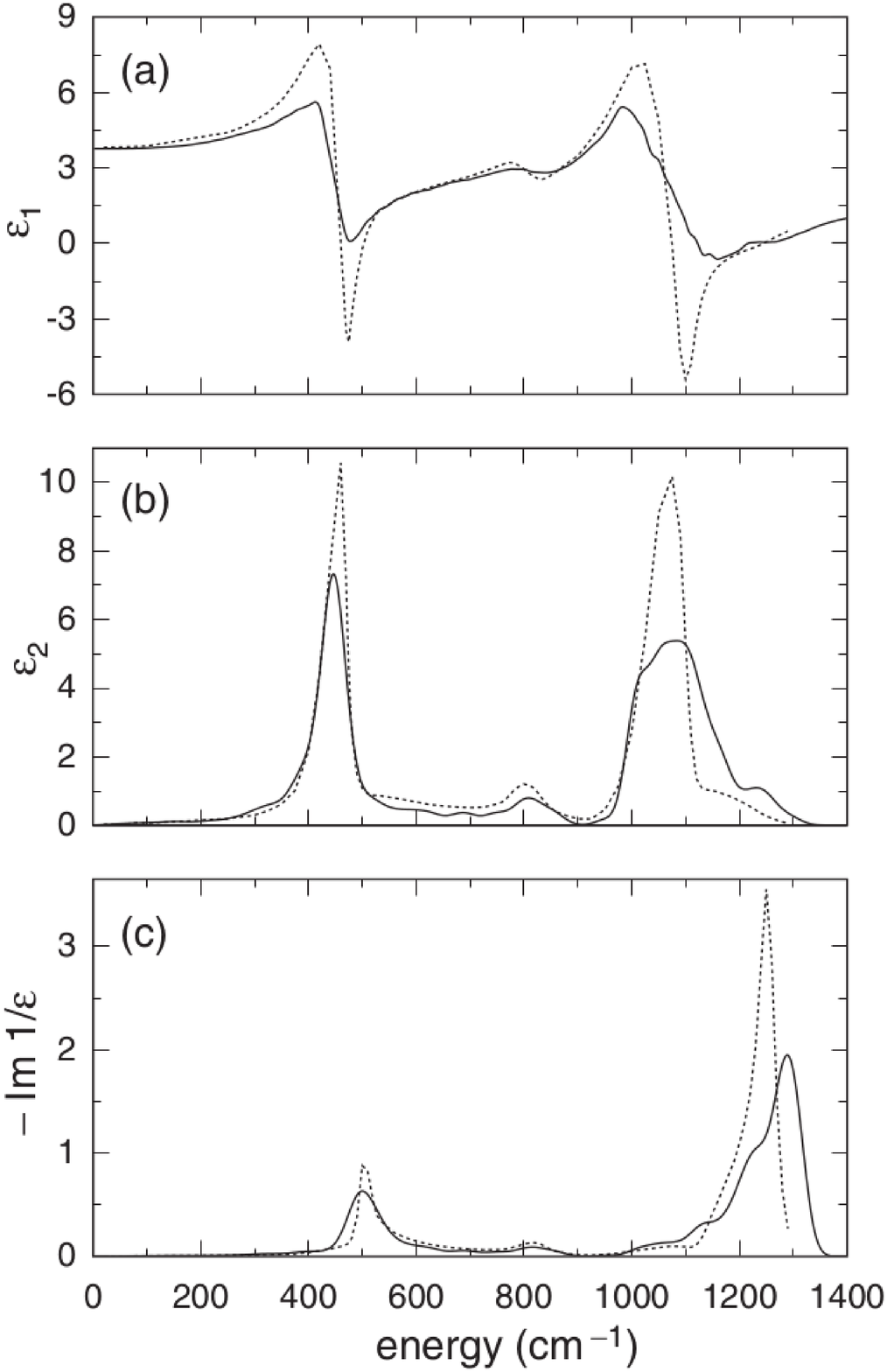}\quad
\includegraphics[scale=0.3]{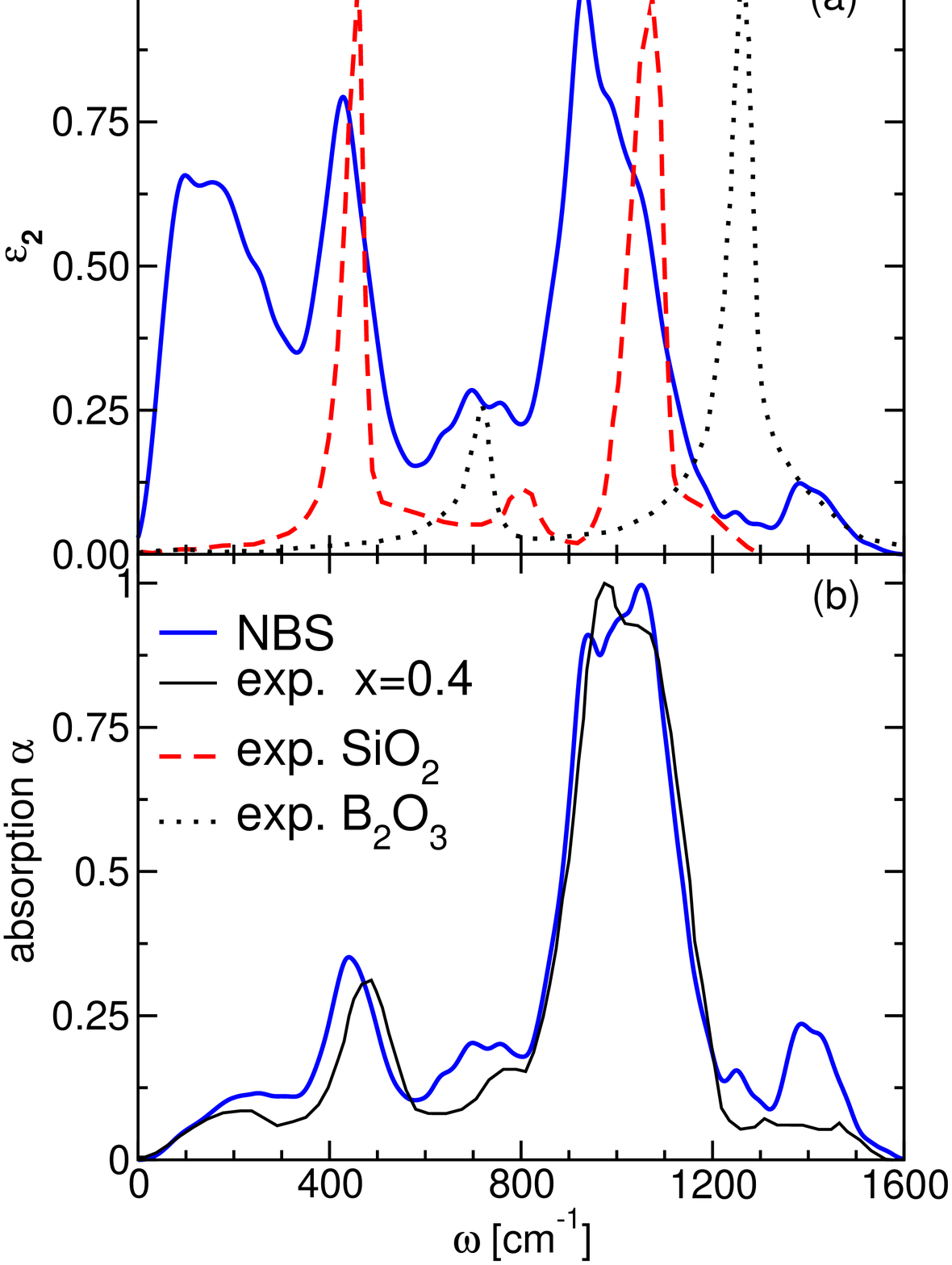}
\caption{Left panel: Infrared spectra for amorphous SiO$_2$ as obtained
from {\it ab initio} simulations (solid lines) and from experiments
(dotted lines)~\cite{giacomazzi2009medium}.  Right panel: Infrared
spectra for a sodium borosilicate glass as obtained from {\it ab initio}
calculations \cite{pedesseau2015first2} and compared to experimental
data for pure SiO$_2$ and B$_2$O$_3$ glasses, as well as for a sodium
borosilicate glass with a similar composition (for more details see
Ref.~\cite{pedesseau2015first2}).}
\label{figure-ir-sio2-giacomazzi2009medium}
\end{figure}

Since all these functions can be measured directly in experiments they
allow on one side to test the predictive power of the simulations
and on the other hand help to interpret the various features found
in the experimental spectra. Examples for this are presented in
Fig.~\ref{figure-ir-sio2-giacomazzi2009medium} where the left panel
shows a comparison between the infra-red spectrum of amorphous silica,
Ref.~\cite{philipp1998silicon}, and the spectrum obtained from an {\it
ab initio} simulation~\cite{giacomazzi2009medium}. One clearly sees that
the theoretical calculation is able to predict reliably the position
of the various peaks, although their width does not match very well
the experimental data~\cite{giacomazzi2009medium}. In the right panel
of the figure we show the corresponding theoretical spectra for a more
complex glass, namely NBS, and compare it with experimental data. From
the upper panel we see that this glass shows below 300~cm$^{-1}$
a broad band. Since in this frequency range the vDOS is strongly
dominated by the motion of Na atoms~\cite{pedesseau2015first2},
we can conclude that this band is related to that atomic species,
which explains why the spectra of the glass-formers that do not
contain alkali atoms do not show such a low frequency band. In addition
the spectra for NBS  shows a narrow band slightly above 400~cm$^{-1}$
. This feature is also present in the spectrum of pure silica and
is known to originate from the bending and rocking motion of oxygen
atoms~\cite{pasquarello1998dynamic,meneses2013investigation}.  Therefore
it is likely that for the NBS glass the origin of the corresponding peak
is the same, although we see that for this glass the peak is somewhat
broader and shifted to lower frequencies. These observations can be
rationalized by the fact that the structure of the NBS glass is more
disordered than the one of pure SiO$_2$.

At high frequencies we can distinguish two bands.  The first one ranges
from 850 to 1200~cm$^{-1}$ , and it can be assigned to oxygen stretching
modes of Si-O bonds~\cite{domine1983study,merzbacher1998structure}. The fact that for
NBS this band is at lower frequencies than the corresponding band for
silica (sharp peak at around 1070~cm$^{-1}$) is consistent with earlier
results which showed that the presence of non-bridging oxygen atoms leads
to a shift of the band to lower frequencies~\cite{ispas2005vibrational}.  The second
high-frequency band, extending between 1200 and 1600 cm$^{-1}$, is due
to the motions of oxygen atoms belonging to $^{[3]}$B, i.e. boron atoms
that are connected to exactly three oxygen atoms.  This is supported by
a comparison with the experimental data for B$_2$O$_3$, which has only
$^{[3]}$B units, and which shows that the latter has a very pronounced
peak in this frequency range.

Finally we show in the lower right panel of
Fig.~\ref{figure-ir-sio2-giacomazzi2009medium} the absorption
spectrum of NBS as obtained from {\it ab initio} simulations as
well as the corresponding experimental data for a very similar glass
composition~\cite{kamitsos1994vibrational}. We see that the theoretical curve
matches very well the experimental data. The main deviation is found
at high frequency in that there the simulation data shows a marked
peak which is absent in the experimental spectrum. The reason for this
difference can be traced back to the fact that the simulation sample has
a too high concentration of $^{[3]}$B units and that these units give
rise to a marked peak in that frequency range~\cite{pedesseau2015first2}.
However, despite this minor flaw we can conclude that the {\it ab
initio} simulation is able to make a surprisingly good prediction of
this quantity.

Finally we mention that DFT based methods have also been proposed
in order to compute Raman and hyper-Raman spectra for periodic
solids \cite{umari2001raman,lazzeri2003first}.  These approaches
have been used to calculate the corresponding spectra for
the main oxide network-formers, i.e. SiO$_2$, B$_2$O$_3$ or
GeO$_2$, and it has been found that they are able to give a good description of the experimental data
\cite{umari2003concentration,giacomazzi2005medium,umari2005fraction,umari2007hyper,ferlat2008boroxol,giacomazzi2009medium}.

\section{NMR spectra}

We conclude with a brief discussion on the calculation of nuclear
magnetic resonance (NMR) spectra within the framework of DFT. Solid
state NMR is a technique that allows to obtain detailed insight into
the local structure of materials (see Chap. XX by J.~F. Stebbins in the present volume
\cite{stebbins2016encyc}). Thus this approach is particularly fruitful
if the system lacks crystalline symmetry, this in contrast to standard
scattering experiments. However, one problem that one faces with NMR
experiments on disordered structures is that sometimes the various
peaks in the measured signal superpose each other and thus it is
not straightforward to come up with a real space interpretation of
the obtained spectrum. It is therefore most useful if a theoretical
calculation can give some guidance for such an interpretation. However,
coming up with a practical computational method that allows to
do such calculations for bulk system is not that easy and in fact
schemes to carried out this kind of calculation were proposed only
15 years ago  \cite{pickard2001all,sebastiani2001new}.  In the
glass community, the so-called GIPAW approach proposed by Pickard
and Mauri \cite{pickard2001all} is certainly the most used one, and
implemented in many {\it ab initio} packages.  However, the theory
on how this is done is somewhat involved and therefore we do not
present it here but instead refer to a recent review on that topic,
see Ref.~\cite{charpentier2011paw}.

\begin{figure}
\includegraphics[scale=0.3]{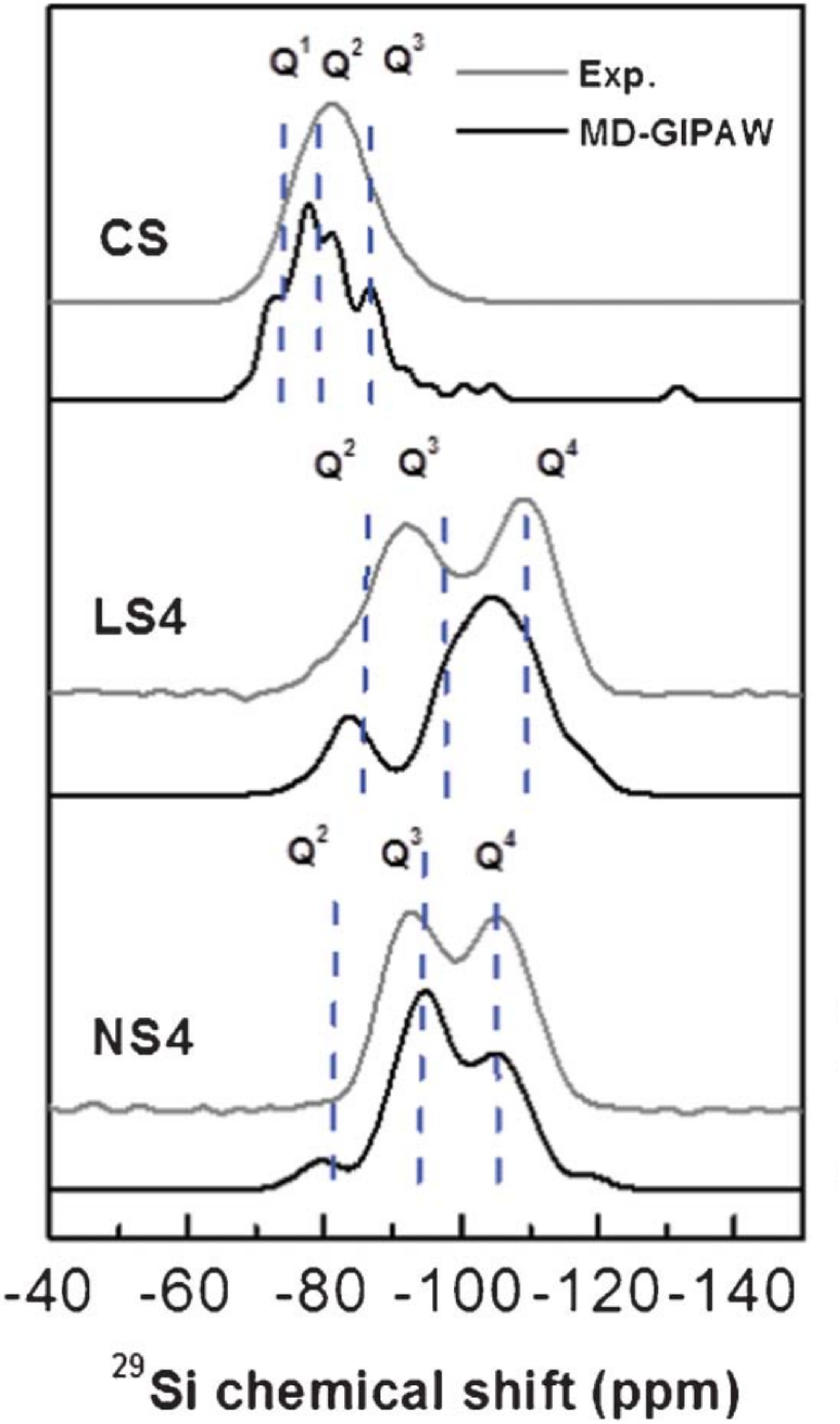}
\includegraphics[scale=0.35]{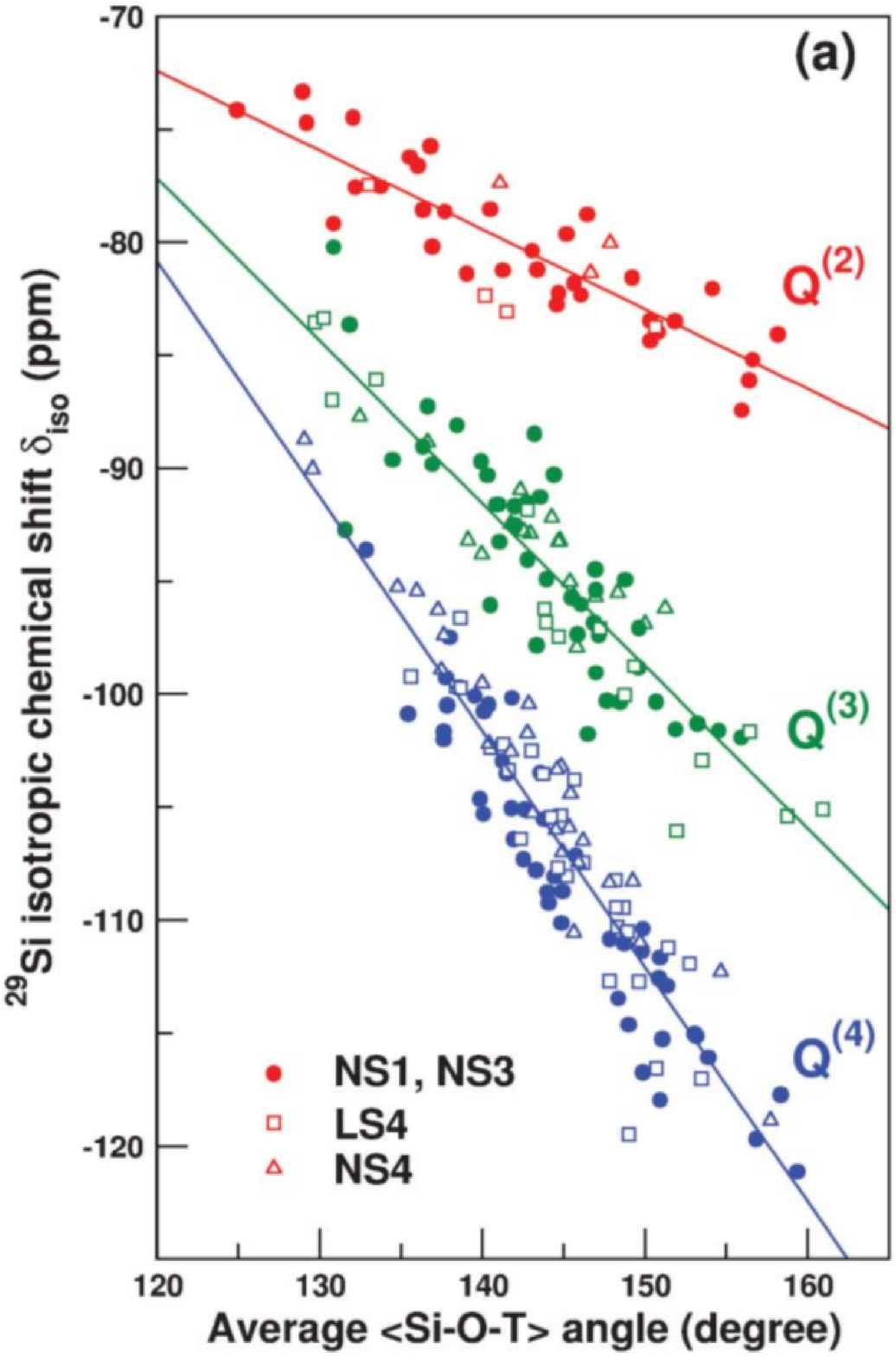}
\caption{
Left panel: Theoretical and experimental $^{[29]}$Si magic angle
spinning NMR spectra of a sodium tetrasilicate glass (molar composition
20Na$_2$O-80 SiO$_2$, labelled NS4), a lithium tetrasilicate glass (molar
composition 20Li$_2$O-80 SiO$_2$, labelled LS4),  and a calcium silicate
glass (50CaO-50SiO$_2$, labelled CS). This panel, extracted from
Ref.~\protect\cite{charpentier2013computational}, compiles data reported
initially in Ref.~\cite{charpentier2004first,ispas2010structural}
for LS4 and NS4 glasses, while the CS spectrum is from
Ref.~\cite{pedone2010multinuclear}.  Right panel: Theoretical prediction
for the variation of the $^{[29]}$Si isotropic chemical shifts as a
function of the mean Si–O–Si bond angle for the different Q$^{(n)}$
species in (a) 43Na$_2$O- 57SiO$_2$ (NS1), 22.5Na$_2$O-77.5SiO$_2$
(NS3), and NS4 and LS4 glasses. The solid lines are linear fits to the
data for the different Q$^{(n)}$ species.  This panel extracted from
Ref.~\protect\cite{charpentier2013computational} compiles data reported
in Ref.~\cite{charpentier2004first,ispas2010structural}
for LS4 and NS4 glasses, while the NS1 and NS3 data were reported in
Ref.~\cite{angeli2011insight}.
}
\label{figure-nmr}
\end{figure}

In the left panel of Fig.~\ref{figure-nmr} we compare for different
glasses experimental NMR data with theoretical prediction. We see
that the agreement is surprisingly good in that the location of
the peaks of the theoretical spectra reproduces well the one of the
experiment. Deviations are sometimes seen in the relative intensity
of the peaks (see, e.g. the case of the lithium-silicate glass). It
is likely that these discrepancies are only partly related  to the
inaccuracy of the theoretical calculation of the spectrum, but instead
also to the too large quench rates used to produce these samples as
well as to their rather modest size. E.g.~in the NBS system discussed
above the concentration of $^{[3]}$B decreases with the cooling rate
whereas the one of $^{[4]}$B increases~\cite{pedesseau2015first2}, which
can be expected to affect the intensity of the various peaks.  However,
with a bit of effort it is possible to handle this problem by probing how
the concentration of the various local structures depend on the cooling
rate (or on the temperature of the liquid) and then to extrapolate these
concentrations to experimental cooling rates~\cite{pedesseau2015first2}.

Although NMR experiments are able to provide information regarding
the nature of the local structure ($Q^{(n)}$-species, connectivity
of the atoms,...) it is very difficult to extract direct information
on the geometry of the local structures, such as bond angles. This
information can of course be obtained directly form the {\it ab initio}
simulations and one finds that the NMR signal is indeed sensitive to these
angles. This is demonstrated in the right panel of Fig.~\ref{figure-nmr}
were we show for various glasses how the chemical shift for an Si atom
depends on the angle formed by Si-O-T, where the O is a first nearest
neighbor oxygen atom of the Si atom and T the second Si atom that is
bonded to that O atom. The data clearly shows that: 1) there is a linear
relation between this angle and the chemical shift; 2) that this relation
is basically independent of the composition of the glass; and 3) that the
slope depends on the $Q^{(n)}$-species of the second Si atom. From these
results one can thus conclude that the NMR spectra do {\it a priori}
contain information not only on the nature of the local structure,
but also on their geometry.

\section{Conclusions and outlook}

In this chapter we have given a brief introduction to the technique
of {\it ab initio} simulations. We have argued that despite the
large computational efforts that it requires, this approach is
very useful and accurate when a multitude of possible local atomic
environnements can coexist, i.e. cases that pose difficulties for
effective potentials. Examples are therefore glasses as well as surfaces,
thus systems which have a large fraction of structural heterogeneities
on the microscopic scale. The goal of this text was not to give an
exhaustive overview of the field of {\it ab initio} simulations since
the relevant literature has already become too extensive.  Therefore we
contented ourself in presenting a few examples that show what current
state of the art simulations of this kind can do. As a consequence many
interesting and important results on a multitude of systems have not been
discussed. Examples are the chalcogenide glasses, i.e.~systems that
have found a widespread application in systems like infrared
lenses, materials for data storage etc$\ldots$ (see Chap. XX by  B. Bureau and J. Lucas in
the present volume~\cite{bureau2016encyc}).  Since these systems
do not contain oxygen, an element that is relatively costly to
simulate via {\it ab initio} simulations, these materials can be
simulated with a {\it relatively} smaller effort per particle and
thus the systems sizes and accessible time scales can become quite
large~\cite{akola2014structure,bouzid2015first}.  As a consequence, and
because of their technological importance, quite a few simulations have
been done on these glass-formers and we refer the reader interested
in more details to the corresponding literature, see for example
Refs.~\cite{blaineau2004electronic,giacomazzi2007first,chaudhuri2009ab,massobrio2009atomic,
lee2010spatial,leroux2013structure,akola2014structure,bauchy2014structural,
micoulaut2014effect,bouzid2015first,raty2015aging}, as well as the
references therein.

A further important topic are glasses that contain or whose surface
is exposed to water. Due to the high reactivity of hydrogen, this
molecule induces local modification of the structure and hence is able
to change the properties of the material even if its concentration is
very small. Effective force fields have difficulties to describe this
reactivity and hence {\it ab initio} simulations are {\it a priori}
much more appropriate. However, these latter type of simulations face
the difficulty that in reality the concentration of water is small,
and hence one needs large systems sizes. As a consequence so far only
relatively few simulations have been done on complex glasses that contain
water~\cite{mischler2002classical,pohlmann2004first,bouyer2010water,tilocca2009modeling,tilocca2010models,
hassanali2010dissociated,tilocca2011initial,spiekermann2012vibrational,berardo2014probing,cimas2014amorphous,spiekermann2016structural}.
We also mention that many first principles investigations have
also been carried out in order to better describe and understand
the properties  of the charged and neutral defects in amorphous
silica, as their presence can strongly affect the performances
of electronic and optical devices based on SiO$_2$
\cite{boero1997structure,bakos2004optically,martin2004oxygen,martin2005neutral,uchino2006density,du2007electron,anderson2011first,giacomazzi2014epr}.

At present the main problem that {\it ab initio} simulations faces is the
large computational effort that it requires. Not only is it expensive
to calculate for a given system the forces acting on a given particle,
but, as mentioned above, this computational work increases quickly with
the number of particles in the system (typically with $N^3$). Thus if
one wants in the future to access significantly larger system sizes
(say $O(10^4)$ atoms in a system with oxygen), one has to avoid this
rapid growth of the computational load. This is exactly the goal of a
some recent new {\it ab initio} techniques  like the
 so-called ``order-$N$'' algorithms
for which the computational cost increase only linearly with the system
size~\cite{bowler2006recent,hine2009linear}, the ``machine learning''
approaches~\cite{caccin2015framework}, or the second generation of the
Car-Parrinello approach~\cite{kuhne2014second}. These methods have been
introduced in the last few years and are promising approaches to allow
accessing in the near future system sizes that are significantly larger
than the ones that have been possible to deal with in the past. Despite
these advances it will in the near future not be possible to simulate
systems with, say, $O(10^5)$ atoms. Systems with such a size or
larger, are, however, important to study, e.g., the mechanical behavior
of glasses on the meso-scale. Thus such simulations will have to be
carried out with effective potentials. But even in this case {\it ab
initio} simulations will be able to play an important role in that they
can be used to develop the needed effective potentials. For example
one can use the detailed microscopic structure information obtained
from an {\it ab initio} simulations of a given material and feed
this information to optimize the parameters present in the effective
potentials~\cite{tangney2002ab,jahn2007modeling,carre2008new,pedone2008ffsioh,kermode2010first,marrocchelli2009construction,izvekov2012mechanism,zeidler2014density,izvekov2015new}.
This approach allows thus to obtain in a systematic manner effective
potentials that are reliable regarding the description of the local
structure but are computationally inexpensive and hence can be used to access
large system sizes or long time scales.

As conclusion we can say that {\it ab initio} simulations of glasses have
become by now a standard technique that allows to measure observables
that are of high practical relevance but inacessable to simulations with
effective potentials. Therefore this approach is, despite the fact that it
is computationally rather costly, a most useful tool to gain insight into
the microscopic properties of glasses and also to develop new type of
glasses with exotic compositions.

\bibliography{q.bib}

\end{document}